\documentclass[
reprint,
nofootinbib,
 amsmath,amssymb,
 aps,
]{revtex4-2}
\usepackage{xcolor}
\usepackage{graphicx}
\usepackage{dcolumn}
\usepackage{bm}
\usepackage{amsmath}
\usepackage{braket}
\usepackage{natbib}
\usepackage[version=4]{mhchem}
\usepackage{gensymb}
\makeatletter
\let\old@makecaption=\@makecaption
\usepackage{subcaption}
\let\@makecaption=\old@makecaption
\makeatother
\usepackage{siunitx}
\usepackage{graphicx,multirow}
\usepackage{array}
\usepackage{hyperref}
\hypersetup{
    colorlinks,
    linkcolor={red!80!black},
    citecolor={blue!80!black},
    urlcolor={blue!80!black}
}
\usepackage[super]{nth}
\usepackage{lipsum}

\newcommand\blfootnote[1]{%
  \begingroup
  \renewcommand\thefootnote{}\footnote{#1}%
  \addtocounter{footnote}{-1}%
  \endgroup
}
\begin{document}

\preprint{APS/123-QED}

\title{Quantum-enhanced quantum Monte Carlo: an industrial view}

\author{Maximilian Amsler$^1$}
\affiliation{QUTAC Material Science Working Group}

\author{Peter Deglmann$^{2,3}$}
\affiliation{QUTAC Material Science Working Group}

\author{Matthias Degroote$^4$}
\affiliation{QUTAC Material Science Working Group}

\author{Michael P. Kaicher$^{3}$}
\affiliation{QUTAC Material Science Working Group}

\author{Matthew Kiser$^{5,6}$}
\affiliation{QUTAC Material Science Working Group}

\author{Michael K{\"u}hn$^{3}$}
\affiliation{QUTAC Material Science Working Group}

\author{Chandan Kumar$^7$}
\affiliation{QUTAC Material Science Working Group}

\author{Andreas Maier$^8$}
\affiliation{QUTAC Material Science Working Group}

\author{Georgy Samsonidze$^9$}
\affiliation{QUTAC Material Science Working Group}

\author{Anna Schroeder$^{10,11}$}
\affiliation{QUTAC Material Science Working Group}

\author{Michael Streif$^{4,\ast}$}
\affiliation{QUTAC Material Science Working Group}

\author{Davide Vodola$^2$}
\affiliation{QUTAC Material Science Working Group}

\author{Christopher Wever$^1$}
\affiliation{QUTAC Material Science Working Group}

\date{\today}

\begin{abstract}
In this work, we test a recently developed method to enhance classical
auxiliary-field quantum Monte Carlo (AFQMC) calculations with quantum computers against examples from chemistry and material science, representatives of classes
of industry-relevant systems.
As molecular test cases, we calculate the energy curve of \ce{H4} and relative energies of ozone and singlet molecular oxygen with respect to triplet molecular oxygen, which are industrially relevant in organic oxidation reactions. We find that trial wave functions beyond single Slater determinants improve the performance of AFQMC and allow to generate energies close to chemical accuracy compared to full configuration interaction (FCI) or experimental results. As a representative for material science we study a quasi-1D Fermi-Hubbard model derived from \ce{CuBr2}, a compound displaying electronic structure properties analogous to cuprates. We find that trial wave functions with both, significantly larger fidelities and lower energies over a Hartree-Fock solution, do not necessarily lead to better AFQMC results.
\end{abstract}

\keywords{QC-QMC, AFQMC, Quantum}

\maketitle

\section{Introduction}
\blfootnote{
$^\ast$ Corresponding author: michael.streif@boehringer-ingelheim.com\\
$^1$ Corporate Sector Research and Advance Engineering, Robert Bosch GmbH, Robert-Bosch-Campus 1, 71272 Renningen, Germany\\
$^2$ BASF SE, Quantum Chemistry, Carl-Bosch-Str. 38, 67063 Ludwigshafen, Germany\\
$^3$ BASF Digital Solutions GmbH, Next Generation Computing, Pfalzgrafenstr. 1, 67056, Ludwigshafen, Germany\\
$^4$ Quantum Lab, Boehringer Ingelheim, Ingelheim am Rhein, Germany\\
$^5$ Volkswagen AG, Ungererstr. 69, 80805 Munich, Germany\\
$^6$ TUM School of Natural Sciences, Technical University of Munich, Boltzmannstr. 10, 85748 Garching, Germany\\
$^7$ BMW Group, New Technology and Innovation, Parkring 19-23, 85748, Garching, Munich, Germany\\
$^8$ Munich Re AG, Munich, Germany\\
$^9$ Robert Bosch LLC, Research and Technology Center, Sunnyvale, CA 94085, USA\\
$^{10}$ Merck KGaA, Frankfurter Stra{\ss}e 250, 64293 Darmstadt, Germany\\
$^{11}$ Quantum Computing Group, Department of Computer Science, Technical University of Darmstadt, Mornewegstra{\ss}e 30, 64293 Darmstadt, Germany}
Recent years have shown significant advancements in the field of quantum computing, both in building more powerful quantum hardware with an ever-increasing number of qubits and lower error rates, as well as in developing quantum algorithms to solve problems in optimization \cite{farhi2014quantum}, machine learning \cite{huang2022provably,huang2022quantum}, and in cryptography \cite{shor1994algorithms}. Finding solutions to classically intractable problems in quantum chemistry and material science is often touted as the first application of future quantum computers in industry \cite{bayerstadler2021industry, santagati2023drug}.

With improvements in quantum hardware and quantum algorithms, the search for areas in industry where quantum computing could provide an economic or technological advantage over current approaches has gathered much attention in the past few years. While in academic studies, much weight is given to demonstrating the superior scaling of a quantum algorithm over a respective classical counterpart in terms of gate complexity in the large-problem-size-limit \cite{Bauer2020,beverland2022assessing}, in an industrial setting, a quantum advantage is reached when the use of a quantum device allows to improve processes, cut down costs, or design new products. Since fully error-corrected quantum computers which can execute quantum algorithms with provable speedups over classical algorithms are years away, it is intriguing to address the question of whether a quantum advantage in Noisy-Intermediate Scale Quantum (NISQ) devices can be found for industrial purposes.

Even though NISQ devices are limited to short quantum circuits, they might provide some advantage over classical algorithms, as computations in classically intractable regions of the Hilbert space are possible even with  modest quantum resources \cite{chen2022complexity,huang2022quantum}. A promising class of NISQ algorithms are variational quantum algorithms, such as the variational quantum eigensolver (VQE) for chemistry problems \cite{peruzzo2014variational,mcclean2016theory}. The VQE is a hybrid algorithm, meaning that the computational task is split between a classical and a quantum processor, while the quantum processor is used only to estimate the energy of a given quantum state manipulated by a set of variational parameters, updated by the classical computer. 
Due to the accumulation of errors with increasing problem size and run time, the largest demonstration of such algorithms has been limited to a few tens of qubits \cite{google2020hartree,huggins2022unbiasing}, even though quantum computers with hundreds of qubits exist \cite{chow2021ibm}. Various classical post-processing steps have been introduced to mitigate the effects of noise \cite{cai2022quantum}. However, the required classical computational overhead resulting from such techniques often nullifies any possible advantage \cite{quek2022exponentially}.   Moreover, the classical optimization of the variational parameters can suffer from vanishing gradients, known as the barren plateau phenomenon \cite{mcclean2018barren}, which can prevent the classical optimization routine from finding the global optimum. In addition, there is some numerical evidence that suggests that the gate-error probabilities needed to generate variational states that describe the ground state of certain molecules within chemical accuracy can lie below the gate-error probabilities required by most quantum error-correction protocols \cite{dalton2022variational}.

The observation of those practical challenges in many numerical simulations and experiments indicates that VQE or related variational algorithms alone are unlikely to generate a quantum advantage in the future. It is therefore important to explore if other algorithms can exploit the computational power present in NISQ devices \cite{arute2019quantum,madsen2022quantum,zhong2020quantum}, and to benchmark the readiness of such new algorithms for industry applications. 

A promising avenue is given by classical post-processing techniques, such as using the output to calculate interaction energies \cite{malone2022towards,loipersberger2022interaction}, improving the energy estimates using neural networks \cite{zhang2022variational}, or in quantum subspace expansions \cite{mcclean2020decoding,klymko2022real,stair2022stochastic}.  In Ref.~\cite{huggins2022unbiasing},  results on \ce{H4} and a small periodic model for the carbon allotrope diamond obtained with a NISQ device suggest that such output can also be used as a trial wave function to guide classical quantum Monte Carlo (QMC) calculations, more precisely auxiliary-field quantum Monte Carlo (AFQMC). 

In this work, we apply AFQMC to a selection of industry-relevant problems. As a molecular test case, we calculate relative energies of ozone and singlet molecular oxygen with respect to triplet molecular oxygen using experimental geometries from Refs.~\cite{tanaka1970coriolis, krupenie1972spectrum} and compare to experimental results \cite{krupenie1972spectrum, weast1983}. To connect to typical industrial workflows, where geometries are optimized with DFT, we compare the AFQMC energies obtained from experimental geometries and from DFT-optimized geometries. As an example from material science, we calculate  the ground state energy of a one-dimensional \ce{CuBr2} chain, mapped to a low-dimensional Hubbard model.
We use trial wave functions obtained from classical methods and VQE circuits. We compare the performance of AFQMC guided by those trial wave functions to mean-field- and other standard quantum chemistry methods. 

From an industrial perspective, it is important to identify the model errors and to determine necessary improvements to a given method.

\paragraph*{About QUTAC:} To investigate the potential impact of quantum computing for industrial applications, thirteen leading German companies are cooperating inside the Quantum Technology \& Applications Consortium (QUTAC). The goal of this collective effort is to evaluate the latest quantum algorithms against industry-relevant applications and provide guidance on needed quantum developments toward industrial applicability.

\section{Auxiliary Field Quantum Monte Carlo (AFQMC) \label{method_AFQMC}}

The AFQMC algorithm is an \textit{ab-initio} method that allows the use of any one-particle basis of size $M$ to project out the ground state of a strongly interacting fermionic system by performing a random walk in the space of fermionic Gaussian states \cite{zhang2003quantum}. Fermionic Gaussian states are exponentials of Hermitian quadratic fermionic operators \cite{bravyi2004lagrangian} and build the basis of AFQMC in both the space of Slater determinants and Hartree-Fock Bogoliubov states \cite{zhang2003quantum,motta2018ab,zhang2013auxiliary,shi2021some,lee2022twenty,shi2017many}. 

Projector methods use the property that the solution of the imaginary time Schr\"odinger equation of the Hamiltonian $H$ asymptotically approaches the ground state $\ket{\Psi_0}$,
\begin{align}
    \ket{\Psi_0} = \lim_{\tau\rightarrow\infty}\ket{\Psi(\tau)} =  \lim_{\tau\rightarrow\infty} \frac{e^{-(H-E_0)\tau}\ket{\Psi_I}}{\sqrt{\braket{\Psi_I|e^{-2(H-E_0)\tau}|\Psi_I}}},\label{mi1}
\end{align}
where $E_0$ is the unknown ground state energy, $\ket{\Psi_I}$ is the initial state, and we assume $\braket{\Psi_I|\Psi_0}\neq 0$. Since $E_0$ is in principle unknown, it is replaced by various adaptive estimators in the AFQMC algorithm \cite{motta2018ab}.  Classical methods have so far not been able to solve Eq.~\eqref{mi1} efficiently for strongly interacting systems, and therefore one has to resort to approximate methods. The core idea of AFQMC is to transform the imaginary time propagator of a quartic operator into an integral over a quadratic operator, whose action on a fermionic Gaussian state can be computed efficiently \cite{Bach1994,kraus2010generalized,shi2017many}. This transformation is realized through a Hubbard-Stratonovich transformation \cite{negele2018quantum}. The resulting integral over matrix exponentials of quadratic operators is then solved in a Monte Carlo fashion~\cite{zhang2003quantum}. 

One hallmark problem of fermionic systems which appears in QMC approaches is the so-called sign or phase problem, which is caused by the anticommutation relations of fermions and phase accumulated in imaginary propagation, respectively \cite{zhang2003quantum,troyer2005computational}. This results in an exponential divergence of the variance of the estimator of the energy in the $k$-th step of a Monte Carlo simulation of $N_\mathrm{W}$ walkers, where the energy estimator is defined as
\begin{align}
    \mathcal E^{(k)}\simeq \frac{\sum_{w=1}^{N_\mathrm{W}}W_{k,w}e^{i\theta_{k,w}}\mathcal E_{\text{loc}}(\Psi_{k,w})}{\sum_{w=1}^{N_\mathrm{W}}W_{k,w}e^{i\theta_{k,w}}},\label{enrg1}
\end{align}
where $W_{k,w}$ and $\theta_{k,w}$ denote the amplitude and phase of the $w$-th walker $\Psi_{k,w}$ at the $k$-th iteration, 
\begin{align}
    \mathcal E_{\text{loc}}(\Psi_{k,w}) = \frac{\braket{\Psi_T|H|\Psi_{k,w}}}{\braket{\Psi_T|\Psi_{k,w}}}\label{enrg2}
\end{align}
is the local energy, and $\ket{\Psi_T}$ is the trial wave function. The latter controls the evolution of the simulation and combined with a phaseless approximation tames the phase problem at the expense of an introduced bias in the energy estimator \cite{zhang2003quantum}.  To be efficiently computable on classical computers, the class of wave functions that can be used as trial states (or walkers) has been limited to linear combinations of (non-orthogonal) Slater determinants or Hartree-Fock Bogoliubov states. However, quantum computers allow us to probe trial wave functions outside this class and thus possibly improve the performance of classical AFQMC calculations, as first realized by Ref.~\cite{huggins2022unbiasing}. In the next section, we describe the classical and quantum trial wave functions used in this work.

\section{AFQMC Trial wave functions}

\subsection{Variational Quantum Eigensolver (VQE)}

The VQE \cite{peruzzo2014variational} is a variational quantum algorithm tailored to find the ground states of a Hamiltonian.
The VQE uses a parameterized quantum circuit or ansatz, and classically optimizes the parameters $\bm{\theta}$ of the circuit with the goal to minimize a cost function, more specifically, the expectation value of the Hamiltonian $\braket{\Psi|H|\Psi}$. It fulfills the variational principle 
\begin{equation}
\label{variation-inequality}
    E = \frac{\braket{\Psi|H|\Psi}}{\braket{\Psi|\Psi}} \geq E_0,
\end{equation} where $E_0$ is the exact ground state energy. Equality holds when $\ket{\Psi} = \ket{\Psi_0}$. 
To keep the number of parameters $\bm{\theta}$ and the depth of the circuit as small as possible, the ansatz is typically tailored specifically to the problem \cite{cerezo2021variational, preskill2018quantum}.

A popular ansatz for chemistry problems is the unitary coupled-cluster ansatz \cite{romero2018strategies}, which we introduce in  Appendix~\ref{app:VQE}. 
Experimental retrieval of the expectation value from quantum hardware can be implemented efficiently by employing schemes such as shadow tomography \cite{hagan2022composite}, basis rotation groupings \cite{huggins2021efficient}, or by optimizing the number of commuting terms \cite{verteletskyi2020measurement}, which reduces the number of required measurements.

The efficient optimization of the parameters $\bm{\theta}$ is an open field of research.  While the well-established gradient-free optimizer COBYLA \cite{powell1994direct} is a popular choice during the prototyping phase, a {\em quantum-aware} optimizer is inevitable to avoid issues associated with noisy quantum hardware. For example, the quantum natural gradient descent \cite{stokes2020quantum} achieves faster convergence by considering the geometric information of quantum states.

\subsection{Matrix product states}
Matrix product states (MPS) provide an efficient parametrization of one-dimensional quantum states in terms of matrices with dimensions bounded by the non-negative integer bond dimension $\chi$~\cite{Orus2014}. The bond dimension can be viewed as a parameter that controls the amount of entanglement and thus the degree of expressivity of an MPS. MPS are variational states that can approximate ground states of a many-body Hamiltonian when used within the density matrix renormalization group (DMRG) method that can capture the entanglement structure of strongly correlated wave functions~\cite{Schollwock2011}.

In this work, we use an MPS representation of the ground state of the Fermi-Hubbard model as a trial wave function $\ket{\Psi_T}$ in AFQMC. We generate  MPS with different bond dimensions $\chi$ by employing the DMRG algorithm implemented in the library ITensor~\cite{itensor}. Choosing different bond dimensions for the MPS provides trial states that approximate the true ground state $\ket{\Psi_0}$ in a controlled manner. This allows us to understand the effect of the trial states on the resulting estimated energy provided by AFQMC.

\section{Results}
\subsection{Molecular systems\label{molecular_systems}}
In the following, we use AFQMC with classical and quantum trial wave functions to calculate (i) the ground state energies of \ce{H4} in a rectangular shape for varying side lengths and  (ii) the relative energies of ozone and singlet molecular oxygen with respect to triplet molecular oxygen. For all AFQMC calculations we use ipie \cite{malone2022a} with computational details summarized in the Appendix in Table~\ref{tab:comp_details}. To generate the Hamiltonian integrals and classical trial wave functions we used PySCF \cite{sun2018pyscf}. To generate the VQE trial wave function, cirq \cite{cirq_developers_2022_7465577} was used. As a quantum trial wave function we use a UCCSD-VQE ansatz applied to a single Slater determinant, see Appendix~\ref{app:VQE} for additional information. To allow for a practical setup, we initialise the VQE parameters with classical CCSD amplitudes \cite{romero2018strategies} and treat the optimisation of the VQE as a black box using scipy's \cite{2020SciPy-NMeth} implementation of the COBYLA optimiser. 

\subsubsection{\ce{H4} Square}

As a first benchmark, we study \ce{H4} in a minimal basis STO-3G  in a rectangular shape with the geometry given by
\begin{align*}
    \ce{H4} :&(\ce{H}, (0, 0, 0)), (\ce{H}, (0, 0, a)),\\ &(\ce{H}, (a, 0, 0)), (\ce{H}, (a, 0, a))\,
\end{align*}
where we vary the side length $a$ from $0.85$\AA{} to $2.5$\AA{} and aim to find the ground state energy of the singlet state. 

The \ce{H4} molecular system is a commonly used benchmark for classical and quantum electronic structure algorithms due to its relatively small size but still present correlation effects \cite{Anderson1979H4,Gasperich2017dqmcH4,huggins2022unbiasing, baek2022say,lee2022twenty}. Due to near degeneracy of the ground state, both the static and dynamic correlations are relevant for an accurate ground state energy calculation \cite{Gasperich2017dqmcH4,Genovese2019dqmcH4}. Moreover, this system was used as the first benchmark in Google's AFQMC landmark paper \cite{huggins2022unbiasing}. Here, we extend this benchmark by calculating the potential energy curve and using a different VQE ansatz. 

We start by running a VQE circuit for each side length using eight qubits. We subsequently use the optimized VQE output as a trial wave function in an AFQMC calculation. 

\begin{figure}[t!]
\centering
\includegraphics[width=1\linewidth]{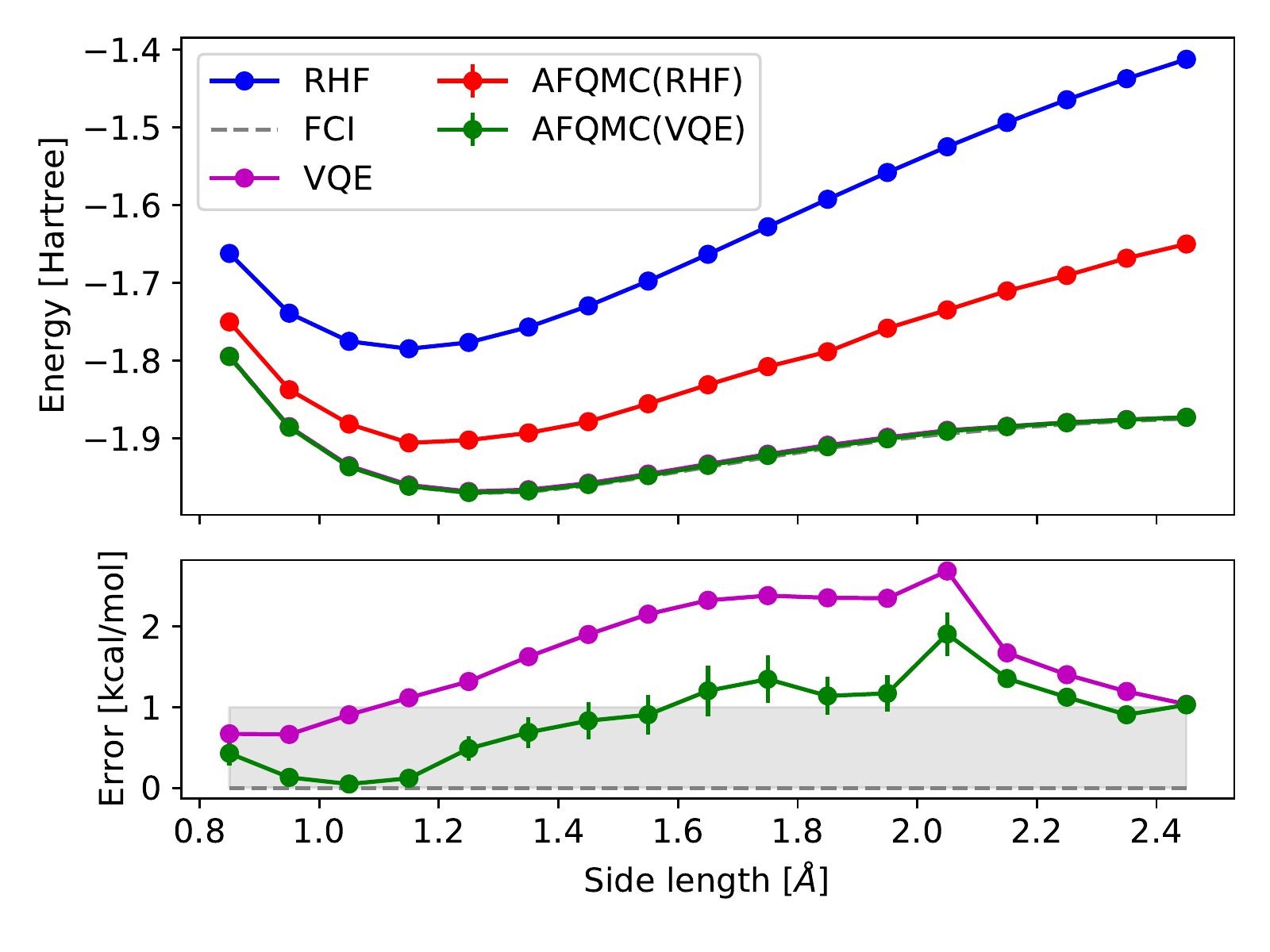}
\caption{(top) Energy surface of the \ce{H4} square with varying side length of the square. For this calculation, a minimal basis STO-3G was used. It should be noted that the FCI and VQE curve are hard to distinguish because of the scale of the y-axis and their similarity to the AFQMC(VQE) results. (bottom) The error of the ground state estimate of plain VQE (purple) and AFQMC(VQE) (green) w.r.t.\ the true ground state energy. The error bars stem from the statistical MC errors.}
\label{fig:h4_curve}
\end{figure}

In Fig.~\ref{fig:h4_curve}(a) we show AFQMC energies together with restricted Hartree-Fock (RHF) and VQE energies and display the exact ground state energy as a reference.  A single-determinant RHF trial wave function is insufficient to guide the AFQMC to the ground state energy of the system \cite{lee2022twenty}. However, the VQE and its AFQMC(VQE) post-processed result can barely be discerned from the exact FCI result in this figure. As shown in Fig.~\ref{fig:h4_curve}(b), VQE alone does not generate results within chemical accuracy for most side lengths. Applying AFQMC as a post-processing procedure to the VQE results improves the energy estimates. However, even for this small example using a noiseless simulation with access to the perfect state vector, AFQMC cannot generate results within chemical accuracy for all side lengths.

\subsubsection{Ozone and molecular oxygen}
\label{sec:relative_energies}

As a second benchmark, we investigate the performance of AFQMC for calculating energy differences between species (relative energies). We calculate the relative energies $\Delta E(x)$ of ozone and singlet molecular oxygen with respect to the triplet ground state of molecular oxygen. While of significant industrial relevance, many standard quantum chemistry methods fail to estimate relative energies and energetic differences between spin states within chemical accuracy. The calculation of the latter with AFQMC was studied in \cite{lee2020utilizing, shee2019singlet}.

In the following, we compare AFQMC results to results obtained from standard quantum chemistry methods and experimental values from \cite{krupenie1972spectrum, weast1983}. We denote the error with respect to the experiment by $\Delta\Delta E(x)=\Delta E(x)-\mathrm{exp. \, value}(x)$. In the case of molecular oxygen respective experimental values are directly available from the literature \cite{krupenie1972spectrum} whereas for ozone the available experimental values refer to enthalpies of formation at absolute zero temperature, $\Delta H^{\mathrm f}_{0K}(\ce{O_3})$, \cite{weast1983} that inherently include a zero-point vibrational energy (ZPVE). To compare our calculated results to those experimental values, we subtract the ZPVE from the enthalpy of formation according to: $\mathrm{exp. \, value}(\ce{O_3}) = \Delta H^{\mathrm f}_{0K}(\ce{O_3}) - \Delta \mathrm{ZPVE}(\ce{O_3}) $ wherein the ZPVEs are calculated at the density functional theory (DFT) level (B3LYP/def2-QZVPP) resulting in $\mathrm{ZPVE}(\ce{^3O_2})=0.0037 \,  \mathrm{Ha}$ and $\mathrm{ZPVE}(\ce{O_3})=0.0064 \, \mathrm{Ha}$ yielding a correction of $\Delta \mathrm{ZPVE}(\ce{O_3}) = 2.2 \,  \mathrm{kJ/mol}$.

Molecular dioxygen \ce{O2}, in particular singlet oxygen \ce{^1O2}, which is formed by electronic excitation of the air constituent triplet oxygen \ce{^3O2}, is a highly reactive molecular species~\cite{Sagadevan2017hydrocarbons, Norbert2008chemrevphotochem} involved in various chemistries such as desired as well as undesired photochemical reactions~\cite{Norbert2008chemrevphotochem,Pibiri2018o2chem,Claude2003singletoxygen}, ene reactions~\cite{Daniel2003Ene-reactions}, and organic oxidation reactions in general. Singlet molecular oxygen is also responsible for damage in biological materials \cite{Triantaphylides2008, Davies2003}.
Ozone (\ce{O3}) is another gaseous, highly reactive form of oxygen that is typically formed in the atmosphere, e.g., by a photochemical reaction catalyzed by \ce{NO_x}. Ozone undergoes a fast chemical reaction with C-C double bond containing materials, such as rubbers, finally cleaving the C-C double bond and is thus responsible for significant material damage  worldwide \cite{Lewis1986}. Due to these reasons - their reactivity and abundance - the modifications of oxygen make for a compelling use-case application of high-level accurate quantum chemical methods. 

Relative energies of singlet molecular oxygen and ozone with respect to triplet molecular oxygen are of particular interest to understand the stabilities of the respective competing species. More generally, an accurate prediction of energy differences, such as relative stabilities, reaction energies, and activation energies, is important in an industrial context as they can be related to experimentally observable thermodynamic and kinetic properties of molecules and chemical reactions.
Previously, there have been theoretical and computational studies on the aforementioned three molecular forms of oxygen \cite{Onishi2018Oxygen,Aleksandr2022CEJ,GRAFENSTEIN1998593,Siebert2001ozone,Holka2010ozone,Dawes2011ozone,Dawes2013ozone} revealing that singlet molecular oxygen and to an even larger extent ozone are challenging systems for single reference electronic structure methods. 

We base our calculations on experimental geometries from Refs.~\cite{tanaka1970coriolis, krupenie1972spectrum}. As a comparison, we use geometries optimized with DFT at the B3LYP/def2-QZVPP level \cite{Becke1993b3lyp, Weigend2005def2bases} resulting in the geometries given in (\ref{eq:dft_geo}). We find that our DFT geometries overestimate the bond length of singlet and triplet molecular oxygen by roughly $0.01$\AA{} and $0.003$\AA{} respectively, with respect to the experiment. For ozone, we find an absolute difference of $0.002$\AA{} in the bond length $r_\mathrm{\ce{O}-\ce{O}}$ and a difference of $0^{\circ}09'$ in the bond angle $\theta_{\ce{O}-\ce{O}-\ce{O}}$. It was furthermore studied in Table~\ref{dft-struct-zpe-comparison} how large the scatter of structural parameters and ZPVEs is when applying different levels of theory at which the computation of energy gradients is well established; here, it turned out, that the differences are small between different classes of density functionals (e.g. ZPVEs vary between 1 and 2 kJ/mol), whereas HF and MP2 predictions are partially far away from DFT and the experimental values.

To have consistent active spaces across all molecules, the molecular orbitals with predominantly atomic $2p$ character are selected, resulting in (8e, 6o) active spaces for the singlet- and triplet molecular oxygen systems and a (12e, 9o) active space for ozone. To further improve the choice of active space orbitals for AFQMC, we run additional CASSCF calculations using the PySCF program package \cite{sun2018pyscf} and base our trial wave function generation on the resulting CASSCF orbitals. We note that such active spaces should be able to capture a large fraction of the correlations present in the system, while still would allow for experimental implementations on currently available quantum hardware. 

To benchmark the performance of AFQMC using a quantum trial wave function, we focus on the ozone system. We first run a UCCSD-VQE (detailed in  Appendix~\ref{app:VQE}) inside the (12e, 9o) active space. The resulting wave function together with the exact solution inside this active space (CAS) and the Hartree-Fock solution (in canonical MOs) is input as trial wave function to benchmark the performance of AFQMC. In Table~\ref{tab:ozone_vqe}, we report the AFQMC results on the experimental geometries of ozone. We find that the energy of AFQMC with a VQE trial wavefunction lies between AFQMC with a HF and a CAS trial wavefunction. 

\begin{table}[]
\begin{tabular}{l|l|l|l}
& HF [Ha]     & VQE [Ha]  & CAS [Ha]   \\ \hline
AFQMC energy & -225.2901(12) & -225.2940(5) & -225.2966(4) 
\end{tabular}
\caption{\label{tab:ozone_vqe}
AFQMC results on the ground state energy of ozone using a HF, VQE and CAS trial wave function. For all calculations cc-pVQZ was used as basis. The VQE and CAS trial wave function were obtained using an (12e, 9o) active space built from CASSCF MOs.}
\end{table}

\begin{figure}[t!]
    \centering
    \includegraphics[width=1\linewidth]{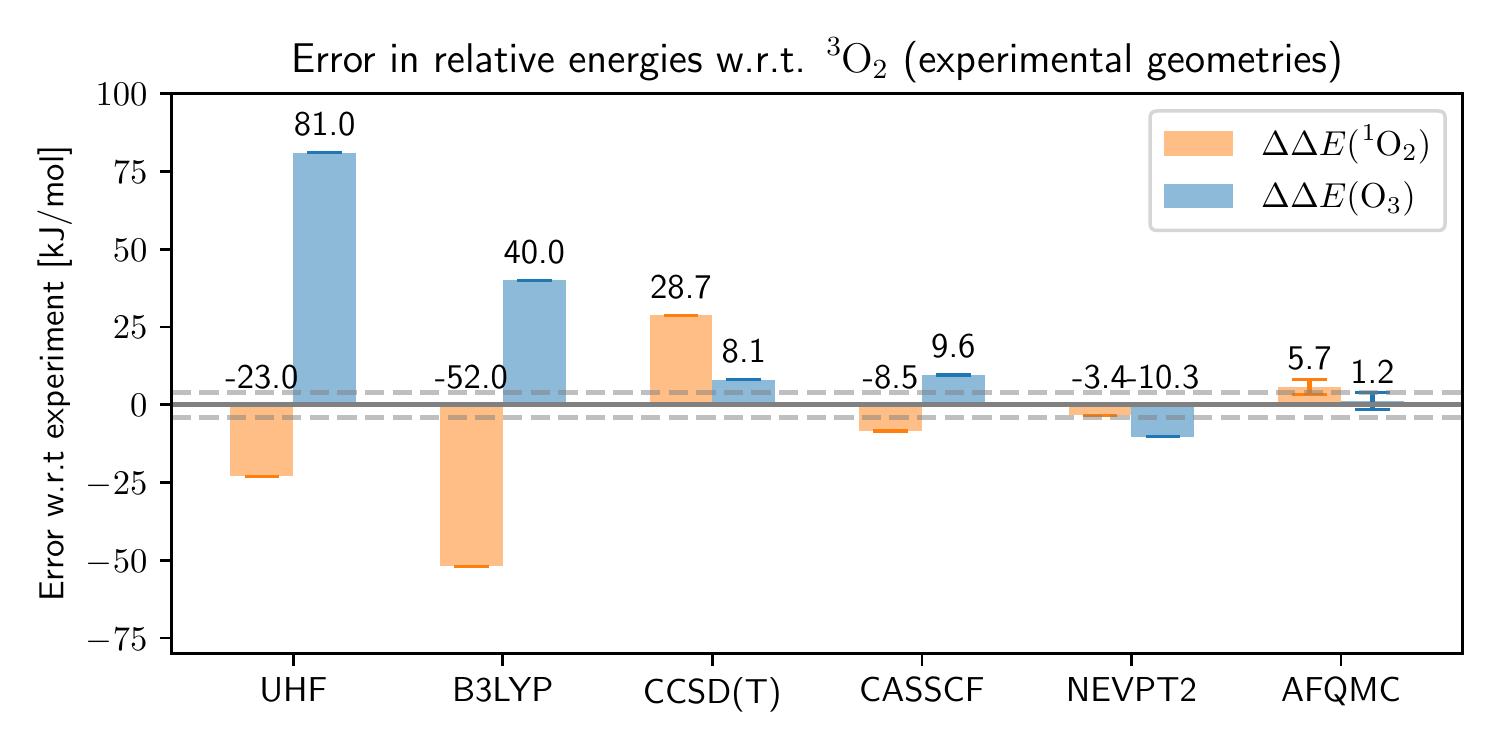}
\caption{Comparison of the error in relative energies of singlet molecular oxygen and ozone with respect to triplet molecular oxygen using experimental geometries given in Eq.~\ref{eq:exp_geo} with respect to experimental data. We report the results from different computational methods obtained using the cc-pVQZ (for UHF) and def2-QZVPP (for B3LYP) basis set and results extrapolated to the CBS limit (otherwise). The CASSCF and NEVPT2 calculations were carried out in a (12e, 9o) active space for ozone  and (8e, 6o) active spaces for singlet and triplet molecular oxygen. The CASSCF wave function was used as a trial wave function in the AFQMC calculation. The error in AFQMC stems from the statistical MC error. 
\label{fig:relative_energies}}
\end{figure}

In the following, we assume that future first-generation quantum computers can generate the exact ground state wave function in a small active space. We use the CAS trial wave functions to benchmark the performance of AFQMC when calculating relative energies. We follow Refs.~\cite{helgaker1997basis, lee2020utilizing} and calculate total energies using the cc-pVTZ and cc-pVQZ basis \cite{Dunning1989basis}. We extrapolate the correlation energies to find the total energies in the complete basis set (CBS) limit, see Appendix~\ref{app:cbs} for more details, and use the CBS total energies to calculate the relative energies. In Fig.~\ref{fig:relative_energies}, we show the results in comparison to the experimental values. We find that while commonly used methods such as DFT (B3LYP), CCSD(T), and CASSCF alone are not able to reach chemical accuracy ($4$ kJ/mol), the AFQMC results, $\Delta E (\ce{^1O_2})=100.4\pm 2.4$ kJ/mol and $\Delta E (\ce{O_3})=144.4\pm 2.7$ kJ/mol, are within or close to chemical accuracy with respect to the experimental results.

However, this comparison has to be taken with a grain of salt. First, for all calculations, the statistical error of AFQMC itself is of the order of a few kJ/mol, making quantitative comparisons between two calculations, which are needed to calculate relative energies, difficult. For calculations of chemical reactions with more than one educt and product, the propagation of errors would get more severe, for example, in redox reactions, $\mathrm{S_N2}$ reactions and many more. 
Second, we only use two points (cc-pVTZ and cc-pVQZ energies) to extrapolate to the CBS limit. Future calculations using larger basis sets (cc-pVXZ with $X\geq 5$) could be used to benchmark the CBS result and to improve it further. Also, the scheme employed here to extrapolate to the CBS limit is one example of many possible choices \cite{helgaker1997basis, feller2011effectiveness}, resulting in a choice-supportive bias. 
Using the same level of theory for calculating the ZPVEs as for calculating the total energies might improve the results in the case of ozone. However, it would not be expected that this would strongly alter the observed trends. This would come at an increased algorithmic cost, as it would require accurate estimating forces and ZPVEs in AFQMC. Lastly, the experimental values themselves are associated with errors. 
We note that, from an industrial perspective, CBS extrapolations are rarely performed due to time constraints. When comparing the plain cc-pVQZ results to the experimental value, we find that the error of AFQMC increases by $4$ kJ/mol, see Table~\ref{tab:results_exp_geo}. When comparing to cc-pVQZ results using DFT-optimized geometries as commonly done in the industry, we find relative energies of $\Delta E (\ce{^1O_2})=104.9\pm 1.2$ kJ/mol and $\Delta E (\ce{O_3})=147.1\pm 1.4$ kJ/mol, comparable to the results obtained with experimental geometries.

\begin{figure*}
\centering
\begin{subfigure}[h]{0.499\textwidth}
    \centering
    \includegraphics[width=1\linewidth]{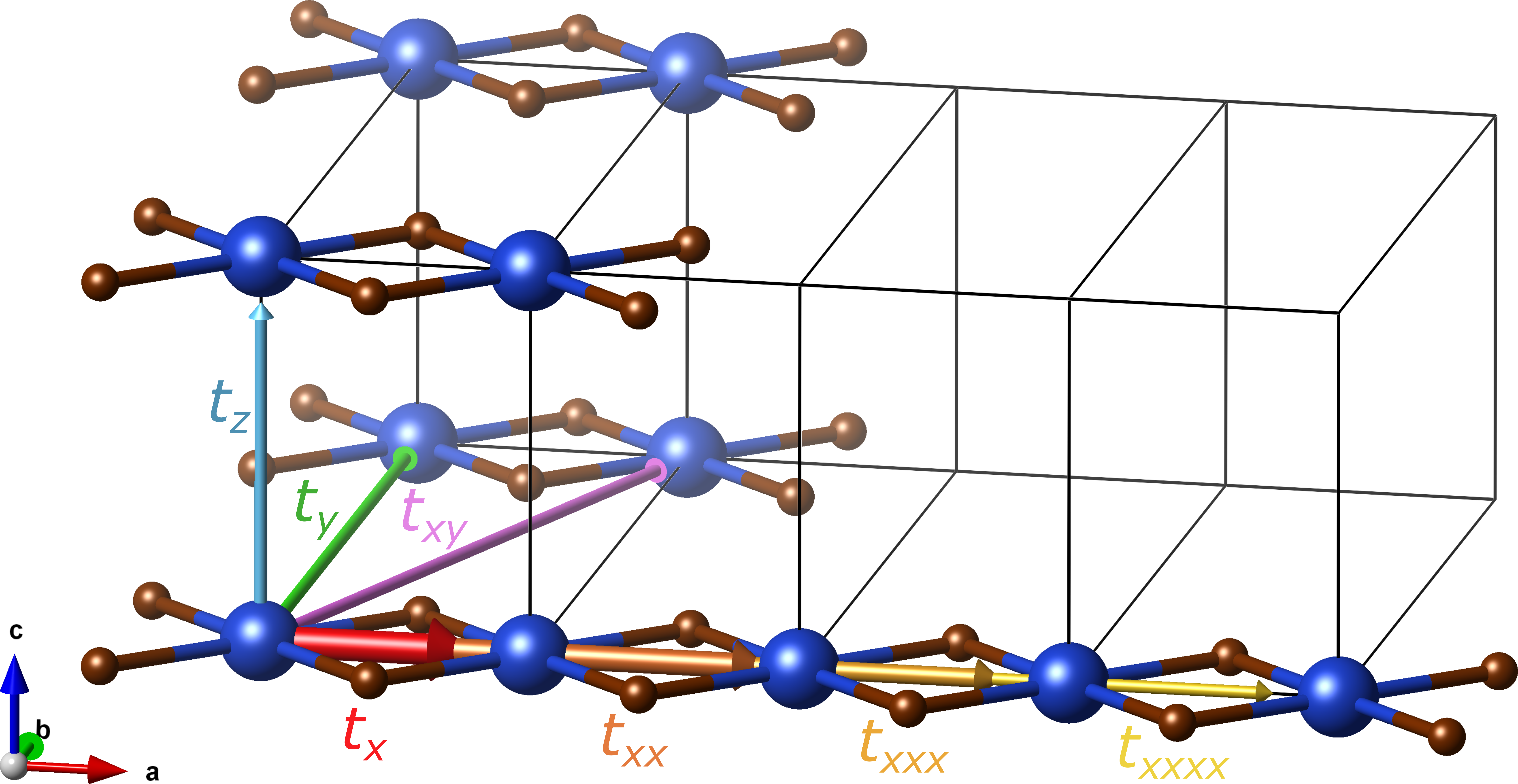}
    \caption{}\label{fig:structural_model_a}
\end{subfigure}%
\begin{subfigure}[h]{0.499\textwidth}
    \centering
    \includegraphics[width=1\linewidth]{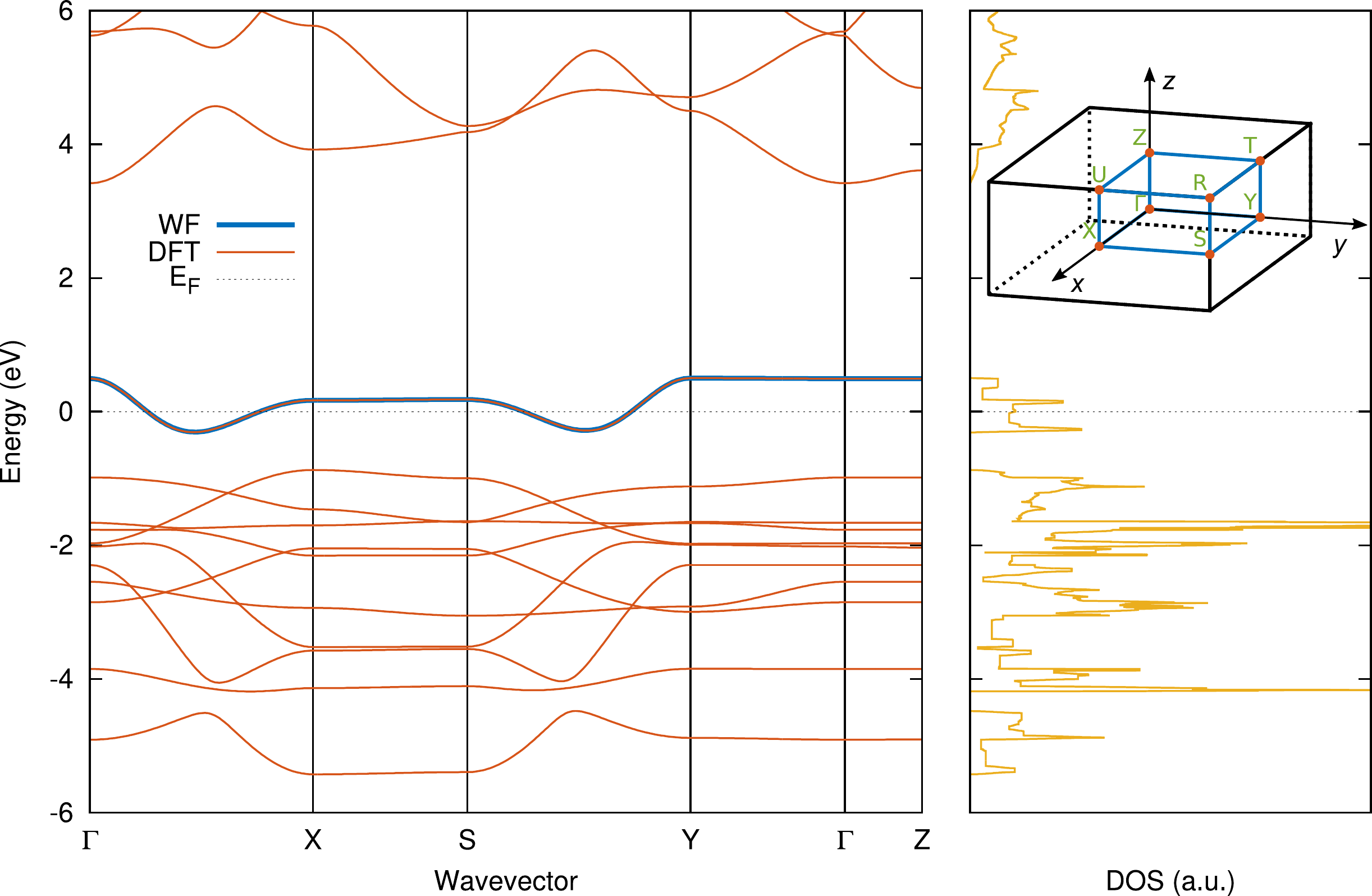}
\caption{}\label{fig:structural_model_b}    
\end{subfigure} 
\caption{(a) Structural model of the quasi-1D chain of \ce{CuBr2}. The blue and brown spheres denote the Cu and Br atoms, respectively, while the periodic images are separated by 10~\AA\ and 5~\AA\ in the lateral (green arrow) and vertical (blue arrow) directions, respectively. The arrows indicate the hopping distance along the chain, ranging from the nearest (red arrow) to the \nth{4} nearest-neighbor (yellow arrow). (b) The band structure of the \ce{CuBr2} chain on the left, together with its density of states (DOS) on the right. Note, that the energy is shifted such that the Fermi level is at value zero. The single wannierized band is shown in blue. The inset shows the Brillouin zone, with the irreducible portion outlined by the blue lines and the labels of the special $k$-points.}
\label{fig:structural_model}
\end{figure*}

\begin{table*}
\definecolor{tx}{HTML}{f81c21}
\definecolor{txx}{HTML}{e47b3e}
\definecolor{txxx}{HTML}{eba838}
\definecolor{txxxx}{HTML}{eed240}
\definecolor{ty}{HTML}{42ae35}
\definecolor{tz}{HTML}{4d90b2}
\definecolor{txy}{HTML}{e385e5}
\begin{ruledtabular}
\begin{tabular}{c c c c c c c c c c}
$\mu=t(000)$ & $t_x=t(100)$ & $t_{xx}=t(200)$ & $t_{xxx}=t(300)$ & $t_{xxxx}=t(400)$& $t_{y}=t(010)$ & $t_{z}=t(001)$ & $t_{xy}=t(110)$ & $U(000, 0)$ & $J(000, 0)$ \\
\hline
$-1.0987$ & 0.0478 & 0.1570 & 0.0339 & 0.0059 & 0.0019 & $-0.0046$ & $-0.0005$ & 4.15 & 4.15
\end{tabular}
\end{ruledtabular}
\caption{Interaction parameters for the Cu:$d_{x^2-y^2}$ orbital of \ce{CuBr2} with respect to lattice vectors $t(xyz)$ in Miller index notation, in eV.\label{tab:cubr2_parameters}}
\end{table*}

\subsection{Extended systems}

\begin{figure*}
\centering
\includegraphics[width=0.9\linewidth]{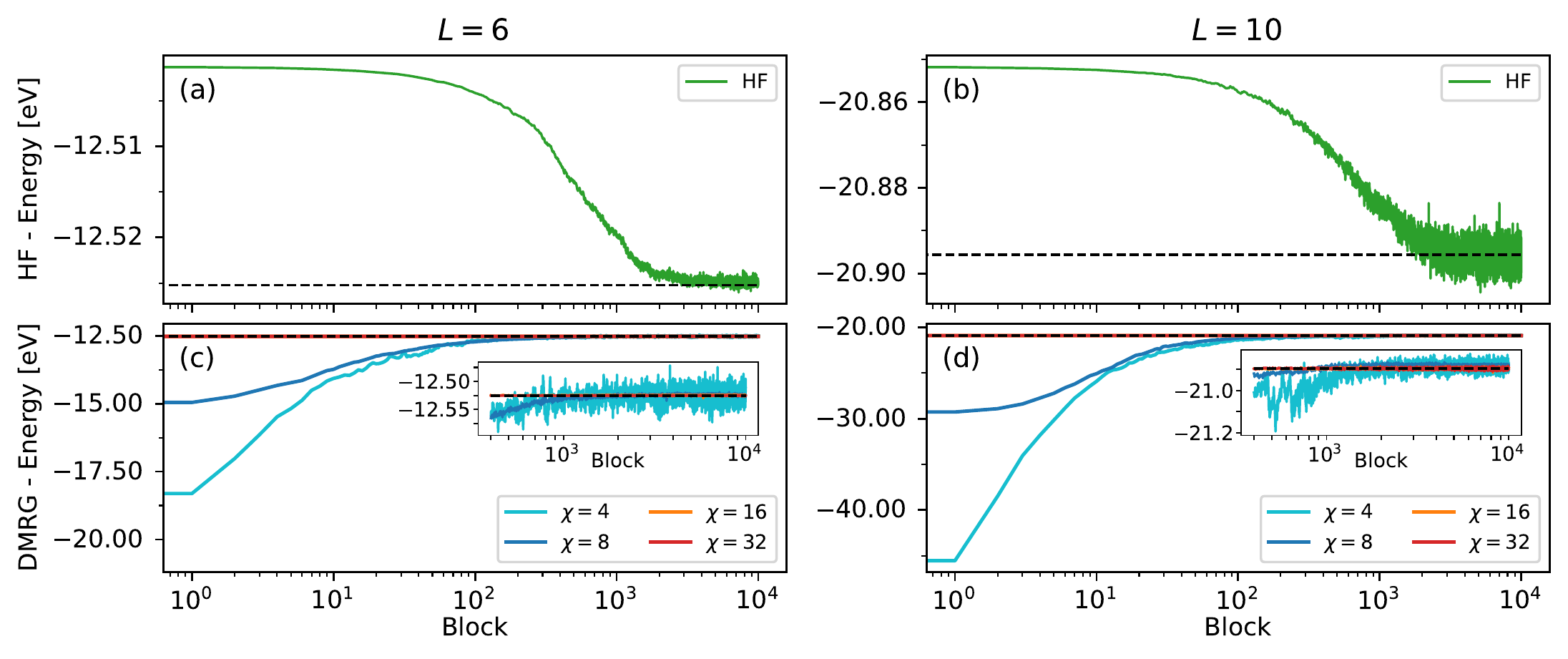}%
\caption{Energy estimator of the Fermi-Hubbard model as a function of the AFQMC projection steps for different trial wave functions. Panels (a) and (b) show the energy estimator obtained from AFQMC using a mean-field HF state as a trial wave function. Panels (c) and (d) show the  results for the energy estimator obtained from AFQMC when an MPS with bond dimension $\chi$ is used as a trial wave function.
\label{fig:AFQMC_hubbard}}
\end{figure*}

Material systems exhibiting effects of strong correlation ranging from metal-insulator transitions to half-metallicity and spin-charge separation are of interest to various technological applications. A particularly interesting class of materials are cuprate high-temperature superconductors, the physics of which are believed to stem from a single correlated $d$ band in the low-energy spectrum~\cite{zhang1988effective, lee2006doping}. A minimal model to describe such a system is the one-band Hubbard model. Still, extensions of it by generalized Hubbard-like Hamiltonian are more refined and, e.g., take explicitly into account the effect of the oxygen $p$-orbitals~\cite{gonzalez_cuprate_1995}. The general atomic structure of cuprates consists of one or several planes of \ce{CuO2} in a square lattice stacked along the $z$-direction, interspersed with layers of guest atoms which act as carrier donors. It is commonly believed that the superconductivity is confined to the 2D-\ce{CuO2} planes and that all relevant spin and charge carriers reside in those substructures~\cite{emery1987theory}.

Analogous to cuprates, copper-bromides consist of similar structural building blocks where \ce{Cu^{2+}} are surrounded by four \ce{Br^{-}} (instead of \ce{O^{2-}} in cuprates) in a square planar arrangement, which in the latter however form chains instead of 2D sheets. Due to this structural and chemical similarity, both materials systems share related electronic properties~\cite{isaacs_materials_2019}, with a single $d_{x^2-y^2}$ band crossing the Fermi level.  Emerging antiferromagnetic~\cite{barraclough_linear_1964,bastow_antiferromagnetism_1981} and multiferroic properties~\cite{zhao_pressure_nodate,zhao_cubr2_2012,zhang_giant_2020} have sparked recent interest in these materials classes, with potential applications in a wide range of magnetoelectric devices.

Native \ce{CuBr2} crystallizes in a polymeric structure with $C2/m$-symmetry, where square planar units of \ce{CuBr4}  (one Cu surrounded by four Br atoms) are linked together to form chains, packed parallel along one direction~\cite{helmholz_crystal_1947}. These chains interact weakly with each other, with the main electronic properties governed by the intra-chain interaction. Here, we map \ce{CuBr2} to an effective, but fictitious 1D model system and compute its properties based on a one-band Hubbard model. We construct the quasi-1D structural model of \ce{CuBr2} by isolating a single chain and placing it in a cell where periodic images of the chains are separated by vacuum of 10 and 5~\AA\ in the lateral and vertical direction, respectively (see Fig.~\ref{fig:structural_model_a}). The band structure and density of states (DOS) are given in Fig.~\ref{fig:structural_model_b}. The Cu--Br distance and Cu--Br--Cu angle within the chain are retained at experimental values of $2.398$\AA{} and $92.35$\degree{}, respectively.

To build the effective model, we use DFT as implemented in the Quantum ESPRESSO package~\cite{giannozzi_quantum_2009} using the Perdew-Burke-Ernzerhof approximation to the exchange-correlation functional~\cite{perdew_generalized_1996} and norm-conserving pseudopotentials~\cite{van_setten_pseudodojo_2018}. We employ a plane-wave cutoff energy of 100~Ry in conjunction with a $k$-points mesh at a density of 0.2/\AA\ to obtain converged results. A single-orbital tight-binding model is constructed by Wannierizing the $d_{x^2-y^2}$ band crossing the Fermi level using the Wannier90 package~\cite{pizzi_wannier90_2020}. The resulting Wannier orbital is shown in Fig.~\ref{fig:cubr2_orbital}, which was converted with wan2respack~\cite{wan2respack} to serve as input for RESPACK~\cite{nakamura-20-respack} to obtain the screened Coulomb interaction parameters based on the constrained random phase approximation (cRPA). We include 83 virtual orbitals in addition to the 17 occupied states, resulting in sufficiently converged screened interaction parameters.

After keeping only the transfer integrals up to the \nth{3} nearest neighbors (see Appendix~\ref{sec:HubbardHamDetails}), the resulting parametrized Hubbard-like Hamiltonian is
\begin{equation}
\begin{split}
    H_{\text{H}} =& -t_x \sum_{i\sigma} a^\dag_{i,\sigma} a_{i+1,\sigma} + \text{H.c.}\\& - t_{xx} \sum_{i,\sigma} a^\dag_{i,\sigma} a_{i+2,\sigma} + \text{H.c.}\\& - t_{xxx} \sum_{i,\sigma} a^\dag_{i,\sigma} a_{i+3,\sigma} + \text{H.c.}\\& - \frac{U}{2} \sum_{i\sigma} n_{i\sigma} + U \sum_i n_{i,\uparrow} n_{i,\downarrow}
    \end{split}\label{ham_bosch}
\end{equation}
where $a^\dag_{i,\sigma}$ ($a_{i,\sigma}$) creates (annihilates) an electron with spin $\sigma$ on the lattice site labelled by $i$ and $n_{i\sigma}=a^\dag_{i,\sigma}a_{i,\sigma}$. The explicit values of the parameters $t_x, t_{xx}, t_{xxx}$ and $U$ (in eV) are reported in Table~\ref{tab:cubr2_parameters}. Note that the spurious hopping terms $t_{xxxx}$, $t_y$, $t_z$, and $t_{xy}$ are small and justify their neglect in our model. Since the ratio $U/\max{(t)}\approx 28\gg 1$, the Hamiltonian~\eqref{ham_bosch} approaches the regime of a spin-1/2 Heisenberg antiferromagnet \cite{auerbach1998interacting,arovas2022hubbard}. The number of lattice sites in the periodic lattice is denoted as $L$, and we use open boundary conditions in the simulations. The Hamiltonian $H_{\text{H}}$ conserves the number of electrons and the total spin, and all simulations are performed at half-filling with balanced spin, $N_\uparrow=N_\downarrow=L/2$. In the following, we study how the fidelity and energy of a trial wave function impact the energy obtained from an AFQMC simulation of Eq.~\eqref{ham_bosch}. 

We compute the ground state energy of $H_{\text{H}}$ via AFQMC using ipie~\cite{malone2022a} with computational details summarized in Table~\ref{tab:comp_details}. We consider two lattice lengths, $L=6$ and $L=10$, and use different wave functions  as trial states $\ket{\Psi_T}$: (i) a Hartree-Fock (HF) mean-field state obtained from an imaginary time evolution of a Slater determinant following the work of \cite{kraus2010generalized}; (ii) a MPS wave function with bond dimension $\chi=4,8,16,32$  obtained via DMRG. For $L= 6$, the MPS at half filling in the zero spin sector generates 400 Slater determinants, all of which are included as a trial wave function. For $L=10$, since including all generated Slater determinants of the MPS state would be too demanding. Instead, we sample Slater determinants with a weight $>5\times 10^{-3}$, resulting in a trial wave function with up to $\sim 300$ Slater determinants. The initial walkers for the AFQMC imaginary time propagation are chosen as the HF state of (i) for both trial states.  

In Fig.~\ref{fig:AFQMC_hubbard} we plot the energy defined in Eq.~\eqref{enrg1} for (a)-(b) a HF trial wave function and (c)-(d) four different MPS states with bond dimensions $\chi=4,8,16,32$ as a function of the block number for $L=6$ (left panels) and $L=10$ (right panels). For the mean-field HF  and the $\chi=4,8$ MPS states, after an initial equilibration phase, the energies reach the exact value. For the higher fidelity MPS with $\chi=16,32$, we do not observe any initial equilibration phase and the local energy oscillates from the beginning around the exact value. This observation can signify that the MPS are already very close to the exact ground state of $H_{\mathrm{H}}$.

\begin{figure}
\centering
\includegraphics[width=0.9\linewidth]{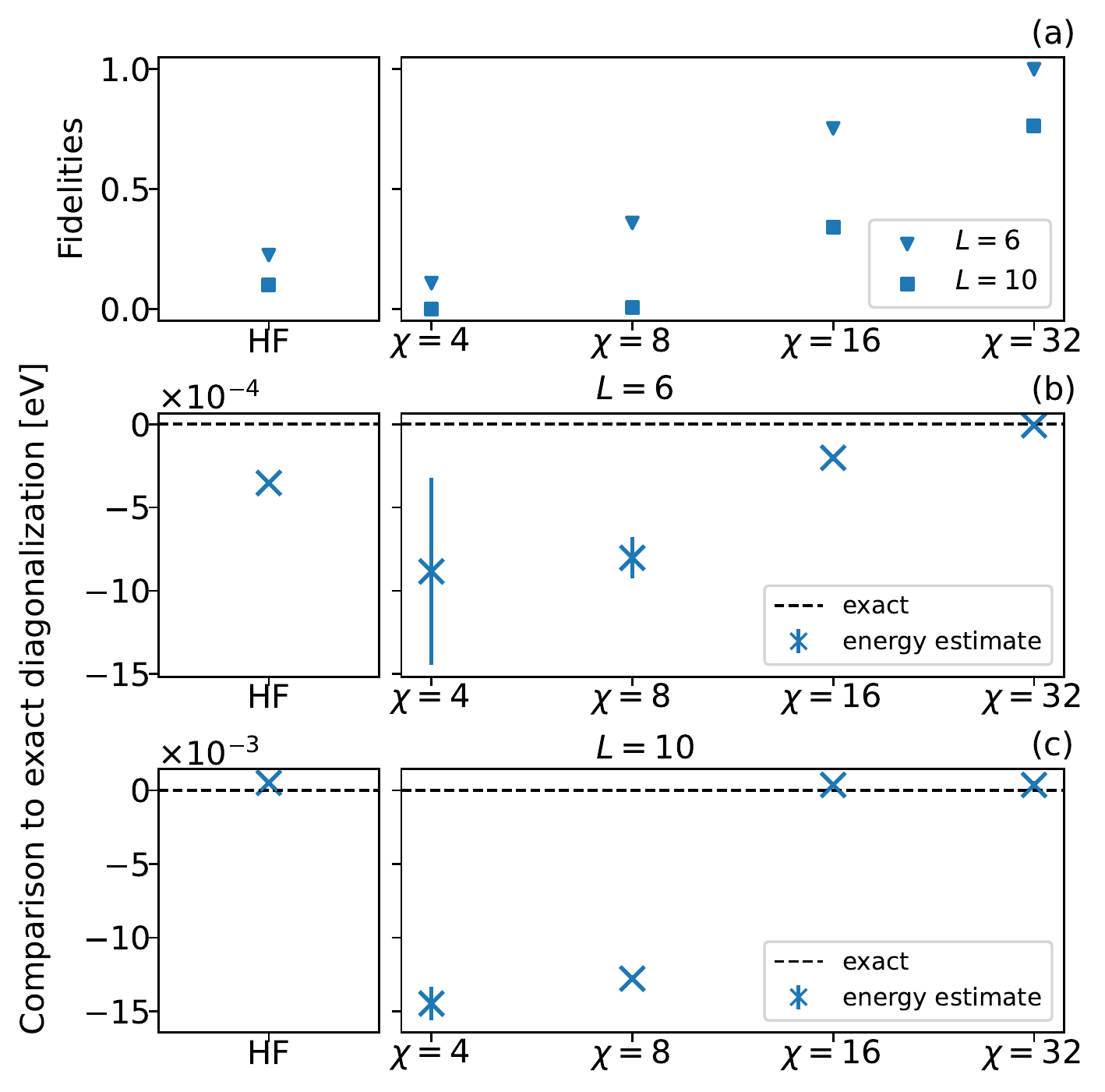}%
\caption{(a) Fidelities of the different trial wave functions with respect to the exact ground state of $H_{\text{H}}$ for two lattice lengths $L=6$ and $L=10$. HF denotes a mean-field state obtained according to \cite{kraus2010generalized}. $\chi = 4,8,16,32$ denote MPS optimized via DMRG with bond dimension $\chi$. 
(b)-(c) Energy estimates from AFQMC for the Hubbard model using different trial wave functions for $L=6$ and $L=10$. The plots show the difference between the converged AFQMC energy averaged over the last 2000 blocks and the energy computed via exact diagonalization. The dashed horizontal line denotes the numerically exact ground state value. See Table~\ref{tab:energy_afqmc_hubbard} for numerical values. \label{fig:AFQMC_energy_diff}}
\end{figure}

To establish a link between the quality of the trial wave function and its effect on the performance of AFQMC, we show the respective fidelities of the trial states and the resulting AFQMC energy estimates in Table~\ref{tab:energy_afqmc_hubbard} and Fig.~\ref{fig:AFQMC_energy_diff}. In Fig.~\ref{fig:AFQMC_energy_diff}(a) we show the fidelities of the different trial states with respect to the exact ground state of $H_{\text{H}}$ for the two lattice lengths $L=6$ and $L=10$. We observe that even though MPS are generally more expressive than a single Slater determinant, one requires a certain amount of entanglement, i.e. sufficiently large bond dimension $\chi$, to improve the fidelity over the HF mean-field solution for the Hamiltonian~\eqref{ham_bosch}. In Fig.~\ref{fig:AFQMC_energy_diff}(b)-(c), we plot the difference of the energy estimator defined in Eq.~\eqref{enrg1} with respect to the true ground state energy $E_0$ for system sizes $L=6$ and $L=10$ and different trial wave functions. 

From Table~\ref{tab:energy_afqmc_hubbard} we find that even though the HF state possesses a lower fidelity than most MPS trial states, its performance in AFQMC is comparable to the performance with a MPS trial state of the largest employed bond dimension. Also, we find that starting with a trial state with a better energy does not guarantee an improved AFQMC result over a trial state with inferior energy. This is an indicator that both the fidelity (or related measures) and energy may not be the only quantities that characterizes the ``goodness'' of a trial wave function in  AFQMC, but that other properties such as symmetry properties of a trial wave function are important, as found in Ref.~\cite{shi2013symmetry,shi2014symmetry}. Such symmetries are present in the HF state, but, generally, not in the state sampled from an MPS output.

\begin{table}
\setlength{\tabcolsep}{3pt}

 \begin{tabular}{|c|c 
S[table-format=1.3] 
S[round-mode=places, round-precision=2,table-format=-1.2e-1]
S[round-mode=places, round-precision=2,table-format=-1.2e-1]|}\hline
 {$L$} & {Trial} & {$|\braket{\Psi_T | \Psi_0}|^2$} & {$E_0 - E_\text{Trial} $ }  & {$E_0 - E_\text{AFQMC}$ }\\\hline
6 &         HF	& 	0.225	& -2.395722e-02	& -3.53e-04\\
&   $\chi=4$	& 	0.108	& -3.700370e-02	& -8.84e-04\\
&   $\chi=8$	& 	0.359	& -9.846918e-03	& -8.02e-04\\
&  $\chi=16$	& 	0.753	& -9.242110e-04	& -2.02e-04\\
&  $\chi=32$	& 	1.000	& -6.890000e-07	& -6.88e-06\\
\hline
10 &         HF	& 	0.101	& -4.365547e-02	& +5.10e-04\\
&   $\chi=4$	& 	0.001	& -8.046097e-02	& -1.45e-02\\
&   $\chi=8$	& 	0.006	& -5.180847e-02	& -1.28e-02\\
&  $\chi=16$	& 	0.340	& -8.892709e-03	& +3.65e-04\\
&  $\chi=32$	& 	0.765	& -8.769410e-04	& +5.97e-04\\
\hline
\end{tabular}
  \caption{Summary of the results for the AFQMC simulations of the Hubbard Hamiltonian $H_\mathrm{H}$ for systems of size $L=6$ and 10. Here, $E_\text{Trial} = \braket{\Psi_T |H_\mathrm{H}|\Psi_T}$ is the expectation value of the energy of the trial wave function, $E_0$ is the exact ground state energy of $H_\mathrm{H}$, $E_\text{AFQMC}$ is the estimate of the energy from the AFQMC. All energies are given in eV. The fidelities (second column) and the energy differences (last column) are also shown in Fig.~\ref{fig:AFQMC_energy_diff}. \label{tab:energy_afqmc_hubbard}}
\end{table}

\section{Conclusion}

In this work, we applied AFQMC with classical and quantum trial wave functions to calculate i) the potential energy curve of \ce{H4}, ii) relative energies of ozone and singlet molecular oxygen with respect to triplet molecular oxygen, and iii) total energies of a \ce{CuBr2} system mapped to a low-dimensional Fermi-Hubbard model.

For \ce{H4}, we found that using a VQE trial wave function (with an energy within a few kcal/mol to the exact ground state energy) does not allow AFQMC to reach chemical accuracy for all side lengths. 
For calculating the total energy of ozone, we considered three different trial wave functions. We found that the AFQMC energies utilizing a VQE trial wave function lie between AFQMC energies using a HF or a CAS trial wave function.
For calculating the relative energies of ozone and singlet molecular oxygen with respect to triplet molecular oxygen, we used AFQMC with CAS trial wave functions, yielding relative energies within or close to chemical accuracy. When using DFT-optimized geometries, commonly used in industry, we found comparable results. 

One source of error is the calculation of the ZPVE, required making the connection with the experimental values. Calculating the zero-point vibrational energy in AFQMC could potentially improve the results. However, this would come at an additional algorithmic cost and make the method less applicable in today's industrial workflows. 

As an example from material science, we provided a non-trivial quasi-1D Fermi-Hubbard Hamiltonian, Eq.~\eqref{ham_bosch}, describing a single chain of \ce{CuBr2}, which exhibits similar electronic structure features as cuprates. We found that na\"ive trial wave functions obtained from sampling from the output of a DMRG calculation that possesses both a better energy and a better fidelity over a simple HF mean-field state do not necessarily lead to better AFQMC results. This suggests that quantum trial wave functions should be physically motivated and respect the expected symmetries of the Hamiltonian. In addition, we find that a simple HF wave function obtained from the method of Ref.~\cite{kraus2010generalized} already provides a trial wave function for which AFQMC can give an energy estimate with an absolute difference of $\sim 10^{-4}$ eV to the FCI results for this system.  Future studies of  \ce{CuBr2} (and related systems)  should follow the strategies of \cite{qin2016benchmark,shi2013symmetry,shi2014symmetry,shi2017many} and investigate the effect of using different Hubbard-Stratonovich transformations, or walkers based on generalized Hartree Fock or Hartree-Fock Bogoliubov states, and the effect of different bases on the AFQMC calculations. Regarding the demands on a quantum algorithm, it will be crucial that a quantum computer can provide a trial wave function that respects the symmetry of the ground state of the problem Hamiltonian. Strategies for designing a quantum state on a quantum computer could follow adiabatic state preparation-inspired variational Ans\"atze of \cite{Wecker_2015}, or center around providing trial states inspired to solve the Heisenberg Hamiltonian in the large$-U$ limit~\cite{auerbach1998interacting}.

In general, a further field of investigation is to understand better the role of the trial wave function with respect to the performance of (classical) AFQMC calculations. Specifically, it remains an open question what properties a trial wave function should possess in AFQMC to generate results within chemical accuracy. For example, whether a high fidelity with respect to the ground state is more important than a low energy. This insight would help to develop specific quantum algorithms tailored to generate good trial wave functions for AFQMC and to understand for which systems a quantum trial wave function would yield an advantage over classically accessible trial wave functions.  Regarding the state preparation of the trial wave function on NISQ hardware, the effect of  noise on AFQMC results remains a topic for further investigation. 

While AFQMC has seen significant improvements over the last two decades \cite{lee2022twenty}, to become a useful tool in the industry, its classical or future quantum-enhanced implementation has to be incorporated into current industrial workflows, made easier to use and show consistent improvement over currently used quantum chemistry methods. The work presented by Ref.~\cite{huggins2022unbiasing, xu2022quantum, yang2021accelerated} and here are the first steps in this direction. 

\begin{acknowledgments}
We thank Fionn Malone and Joonho Lee for insightful discussions and early access to the AFQMC python package ipie \cite{malone2022a}. MA, CW, and GS thank Takashi Koretsune and Kazuma Nakamura for fruitful discussions concerning cRPA calculations. We thank Joonho Lee, Nikolaj Moll and Clemens Utschig-Utschig for their feedback on the manuscript.
\end{acknowledgments}

\bibliographystyle{apsrev4-2}
\bibliography{new_afqmc}
\newpage
\appendix
\onecolumngrid

\section{VQE UCCSD ansatz}
\label{app:VQE}
The UCC-VQE ansatz \cite{romero2018strategies} originated from the Coupled Cluster theory \cite{bartlett2005theory}, where the wave function is represented as
\begin{equation} \label{eq:CC}
    \ket{\Psi} = e^{T} \ket{\mathrm{HF}},
\end{equation}
with $T$ being the excitation operator given by
\begin{gather}
    T = \sum^{\eta}_{i = 1} T_i, \\
     T_1 = \sum_{\substack{i\in \mathrm{occ} \\ a \in \mathrm{virt}}} t^i_a a_a^\dagger a_i, \\
         T_2 = \sum_{\substack{i>j\in \mathrm{occ} \\ a>b \in \mathrm{virt}}} t^{ij}_{ab} a_a^\dagger a_b^\dagger a_i a_j\, ,\\
         \dots
\end{gather}  
Depending on the truncation order of the series, the CC method is called CCSD (single and double excitation) or CCSDT (single, double, and triple), and so forth.  
A unitary version of Eq.~(\ref{eq:CC})  ~\cite{taube2006new}
\begin{equation}
    \ket{\Psi} = e^{T - T^\dagger} \ket{\mathrm{HF}}, \quad 
\end{equation} allows serving as a VQE ansatz on a quantum computer, where the  amplitudes $t$ are variational parameters. To implement the unitary on a quantum computer, first-order Trotterization formulas \cite{trotter1959product,suzuki1976relationship} can be used, resulting in 
\begin{equation}
    U(\bm{t}) \approx U_{\mathrm{Trott}} (\bm{t}) = \prod_i e^{t_i(\tau_i - \tau_i^\dagger)}\,.
\end{equation}
Bravyi-Kitaev or Jordan Wigner transformations \cite{bravyi2002fermionic}  convert the operators $\tau_i$ into products of Pauli gates native to quantum hardware. One aspect that should be mentioned regarding the practical application of the UCCSD ansatz is the parameter initialization. While the VQE optimization usually converges even with the Hartree-Fock state as the starting point (i.e., initializing all parameters with zero), one can significantly reduce the number of training iterations by taking results from perturbation methods, e.g., MP2 \cite{head1988mp2}, or CCSD calculation as the initial parameter guess. Recent findings suggest that initialization of the parameters through CCSD leads to significantly better results than initializing through MP2~\cite{hirsbrunner2023beyond}.
 

\section{Molecular structures \label{app:geometries}}
For the calculations on singlet and triplet molecular oxygen, and ozone presented in Sec.~\ref{sec:relative_energies}, Table~\ref{tab:results_exp_geo} and Table~\ref{tab:results_dft_geo}, we use experimental geometries from \cite{tanaka1970coriolis, krupenie1972spectrum}:
\begin{gather}
\ce{^3O2}:
\begin{tabular}{|c| S[table-format=+1.7] | S[table-format=+1.7] |S[table-format=+1.7]|}\hline
{} & {$x$/\AA{}} & {$y$/\AA{}} & {$z$/\AA{}} \\ \hline
\ce{O} & 0.0 & 0.0 &  0.603760\\
\ce{O} & 0.0 & 0.0 & -0.603760\\ \hline
\end{tabular} \nonumber \\
\ce{^1O2}:
\begin{tabular}{|c| S[table-format=+1.7] | S[table-format=+1.7] |S[table-format=+1.7]|}\hline
\ce{O} & 0.0 & 0.0 &  0.607800\\
\ce{O} & 0.0 & 0.0 & -0.607800\\ \hline
\end{tabular} \label{eq:exp_geo}  \\
\ce{^{\phantom{3}}O3}:
\begin{tabular}{|c| S[table-format=+1.7] | S[table-format=+1.7] |S[table-format=+1.7]|}\hline
\ce{O} & 0.0 & 0.0 &  0.0\\
\ce{O} & 0.0 & 0.0 &  1.2717000\\ 
\ce{O} & 1.1383850 & 0.0 & 1.8385340\\ \hline
\end{tabular}\nonumber 
\end{gather}

as well as geometries optimized at the density functional theory (DFT) level (UB3LYP/def2-QZVPP), resulting in the xyz-structures:  
\begin{gather}
\ce{^3O2}:
\begin{tabular}{|c| S[table-format=+1.7] | S[table-format=+1.7] |S[table-format=+1.7]|}\hline
{} & {$x$/\AA{}} & {$y$/\AA{}} & {$z$/\AA{}} \\ \hline
\ce{O} & 0.0 & 0.0 &  0.6020120\\
\ce{O} & 0.0 & 0.0 & -0.6020120\\ \hline
\end{tabular} \nonumber \\
\ce{^1O2}:
\begin{tabular}{|c| S[table-format=+1.7] | S[table-format=+1.7] |S[table-format=+1.7]|}\hline
\ce{O} & 0.0 & 0.0 &  0.6021008\\
\ce{O} & 0.0 & 0.0 & -0.6021008\\ \hline
\end{tabular} \label{eq:dft_geo} \\
\ce{^{\phantom{3}}O3}:
\begin{tabular}{|c| S[table-format=+1.7] | S[table-format=+1.7] |S[table-format=+1.7]|}\hline
\ce{O} &  3.1946229 & 3.9056962 &  0.0\\
\ce{O} & -2.1046353 & 3.2467339 &  0.0\\ 
\ce{O} & -1.0278818 & 3.9270800 & 0.0\\ \hline
\end{tabular}\nonumber
\end{gather}

\begin{table}
\begin{tabular}{c|c|c|c|c|c|c} 
  &  & HF & MP2 & BP86 & B3LYP & M06-2X \\ \hline
 \ce{^3O2} & reference  & UHF & UHF & UKS & UKS & UKS \\
 & $r_\mathrm{\ce{O}-\ce{O}}$ [pm] & 115.79 & 121.90 & 122.03 & 120.43 & 118.79 \\
 & ZPVE [kJ/mol] & 11.78 & 8.84 & 9.24 & 9.78 & 10.55 \\ \hline
  \ce{^1O2} & reference  & UHF & RHF & UKS & UKS & UKS \\
 & $r_\mathrm{\ce{O}-\ce{O}}$ [pm] & 115.61 & 124.34 & 122.12 & 120.46 & 118.73 \\ 
 & ZPVE [kJ/mol] & 11.80 & 7.72 & 9.20 & 9.75 & 10.54 \\ \hline
\ce{O3} &  reference  & UHF & RHF & RKS (=UKS) & UKS & UKS \\
 & $r_\mathrm{\ce{O}-\ce{O}}$ [pm] & 127.54 & 127.81 & 127.55 & 127.41 & 125.69 \\
 & $\theta_{\ce{O}-\ce{O}-\ce{O}}$ [$^{\circ}$] & 118.14 & 116.76 & 118.14 & 116.55 & 116.05 \\
 & ZPVE [kJ/mol] & 17.66 & 24.69 & 17.63 & 16.87 & 18.57 \\ 
\end{tabular} 
\caption{A benchmark on the method dependence of structural parameters as well as zero-point vibrational energies (ZPVE). All calculations are performed using the cc-pVQZ basis.\label{dft-struct-zpe-comparison}}
 \end{table}
\section{Complete Basis Set (CBS) Extrapolation}\label{app:cbs}

To extrapolate the total energies of the singlet and triplet molecular oxygen, and ozone, we follow \cite{helgaker1997basis, halkier1998basis}. We first subtract the RHF/R(O)HF energies from the results of the correlated methods energies (AFQMC(CAS), CASSCF, CCSD(T) and NEVPT2) to extrapolate the correlation energy using the function \cite{helgaker1997basis}
\begin{align}
    E_\mathrm{corr}^X = E_\mathrm{corr}^\infty+a X^{-3}\,,   
\end{align}
with $a$ being a free parameter and $X$ denoting the cardinal number (in cc-pV$X$Z). As in Ref.~\cite{lee2020utilizing}, we only use the total energies from $X=\{T,Q\}$ together with the cc-pV5Z RHF/R(O)HF energy as CBS Hartree Fock energy $E_{HF}^\infty$.

\section{Hubbard model Hamiltonian}\label{sec:HubbardHamDetails}
The Hamiltonian of the generalized (multi-orbital) Fermi-Hubbard model in the second-quantization form is given by~\cite{nakamura-20-respack, charlebois-21-hubbard-model, pavarini-2011-strongly-correlated-materials, lechermann-06-hubbard-model, castellani-78-hubbard-model, fresard-97-hubbard-model, imada-98-hubbard-model}:
\begin{align}
    H = \sum_{ijmm'\sigma} t_{mm'} \left( \mathbf{R}_j - \mathbf{R}_i \right) a_{im\sigma}^{\dagger} a_{jm'\sigma} + \sum_{imm'\sigma\sigma'} \bigl( U_{mm'} \left( \mathbf{0}, 0 \right) - J_{mm'} \left( \mathbf{0}, 0 \right) \delta_{\sigma\sigma'} \bigr) n_{im\sigma} n_{im'\sigma'},
    \label{eq:fh}
\end{align}
where $i$ and $j$ are lattice indices, $\mathbf{R}_i$ and $\mathbf{R}_j$ are lattice vectors, $m$ and $m'$ are orbital indices, $\sigma$ and $\sigma'$ are spin indices, $a_{im\sigma}^{\dagger}$ and $a_{im\sigma}$ are creation and annihilation operators, and $n_{im\sigma} = a_{im\sigma}^{\dagger} a_{im\sigma}$ is the particle number operator. The transfer integral and direct and exchange Coulomb integrals are given by:
\begin{gather}
    t_{mm'} \left( \mathbf{R}_j - \mathbf{R}_i \right) = \int \phi_{im}^\ast \left( \mathbf{r} \right) H_\mathrm{MF} \left( \mathbf{r} \right) \phi_{jm'} \left( \mathbf{r} \right) \,d\mathbf{r}, \\
    U_{mm'} \left( \mathbf{R}_j - \mathbf{R}_i, \omega \right) = \iint \phi_{im}^\ast \left( \mathbf{r} \right) \phi_{im} \left( \mathbf{r} \right) W \left( \mathbf{r}, \mathbf{r}', \omega \right) \phi_{jm'}^\ast \left( \mathbf{r}' \right) \phi_{jm'} \left( \mathbf{r}' \right) \,d\mathbf{r}\,d\mathbf{r}', \\
    J_{mm'} \left( \mathbf{R}_j - \mathbf{R}_i, \omega \right) = \iint \phi_{im}^\ast \left( \mathbf{r} \right) \phi_{jm'} \left( \mathbf{r} \right) W \left( \mathbf{r}, \mathbf{r}', \omega \right) \phi_{jm'}^\ast \left( \mathbf{r}' \right) \phi_{im} \left( \mathbf{r}' \right) \,d\mathbf{r}\,d\mathbf{r}',
\end{gather}
where $\mathbf{r}$ and $\mathbf{r}'$ are spatial coordinates, $\omega$ is frequency, $\phi_{im}$ are Wannier orbitals, $H_\mathrm{MF}$ is the mean-field Hamiltonian, and $W$ is the screened Coulomb interaction potential calculated within the constrained random phase approximation (cRPA)~\cite{nakamura-20-respack}. Note that $t_{mm} \left( \mathbf{0} \right)$ are the orbital energies, which can be omitted in the case of degenerate orbitals. Also note that $U_{mm} \left( \mathbf{0}, \omega \right) = J_{mm} \left( \mathbf{0}, \omega \right)$, which ensures the Pauli exclusion principle in Eq.~\eqref{eq:fh}.

\begin{figure}
    \begin{subfigure}[b]{0.48\textwidth}
    \centering
    \includegraphics[width=.95\linewidth]{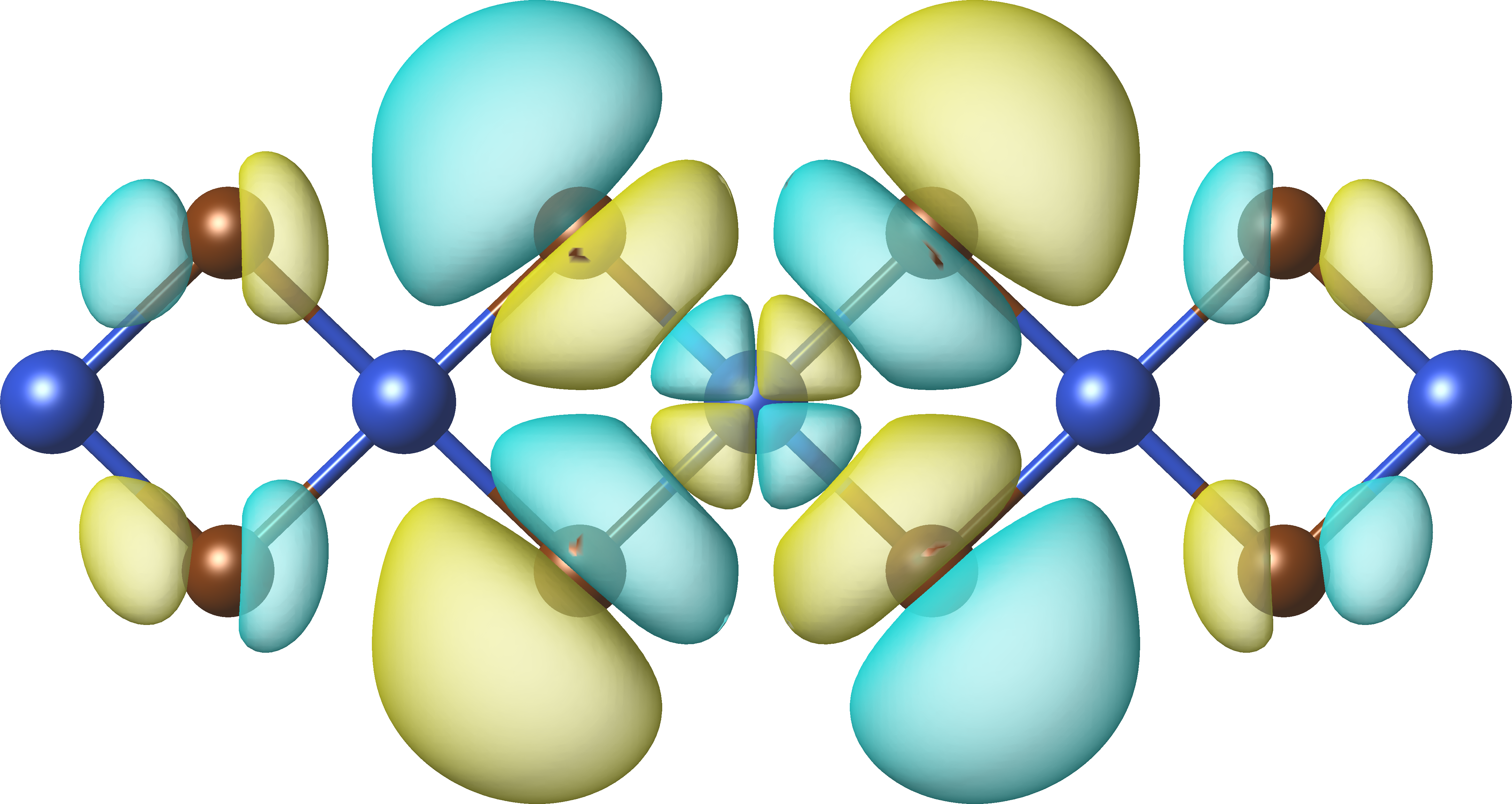}
    \caption{}
    \label{fig:cubr2_orbital}
    \end{subfigure}
    \begin{subfigure}[b]{0.48\textwidth}
     \centering
    \includegraphics[width=.95\linewidth]{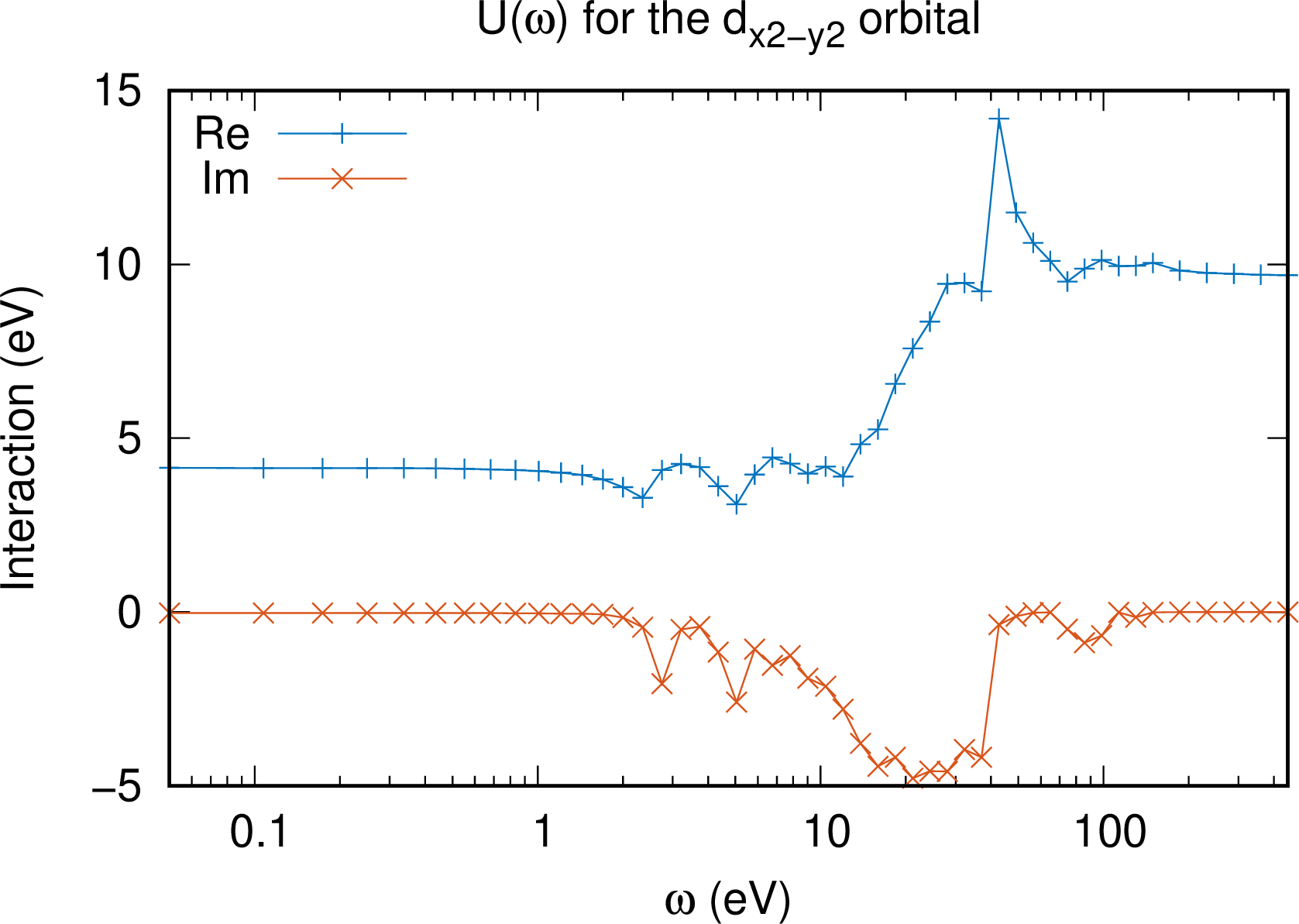}
   \caption{}
    \label{fig:cubr2_u}
    \end{subfigure}
    \caption{(a) Isosurface plot of the Wannier orbital of \ce{CuBr2} with $d_{x^2-y^2}$ character, centered on a Cu atom (b) Frequency-dependent direct Coulomb integral for the Cu:$d_{x^2-y^2}$. The value of $U_\text{H}$ for our Hubbard model was taken as the static limit of the real part of $U(\omega)$, i.e., $U_\text{H}=\lim \limits_{\omega \to 0}\Re\left[U(\omega)\right]$.}
    \label{fig:cubr2_orbital_and_cubr2_u}
\end{figure}

Based on the Hamiltonian of the generalized (multi-orbital) Fermi-Hubbard model, the Wannier orbital and the frequency-dependent direct Coulomb integral plots of the Cu:$d_{x^2-y^2}$ orbital of \ce{CuBr2} are shown in Fig.~\ref{fig:cubr2_orbital} and \ref{fig:cubr2_u}, respectively, and the interaction parameters are given in Table~\ref{tab:cubr2_parameters}. A comparison of these values suggests that transfer integrals up to the \nth{3} nearest neighbor (i.e., $t_{x}$, $t_{xx}$, and $t_{xxx}$) must be included in the model. The terms $t_{xxxx}$, $t_y$, $t_z$, and $t_{xy}$ are small and are thus neglected.

\begin{table}[!]
\begin{tabular}{l|l|l|l}                  & \ce{H4}                    & \ce{O2} \& \ce{O3}              & Hubbard system (\ce{CuBr2}) \\ \hline
Number of walkers $N_\mathrm{W}$            & $10^3$ & $10^3$ & $10^3$                       \\ \hline
Number of blocks  $N_\mathrm{B}$             & $10^3$ & $10^4$ & $10^4$                \\ \hline
Time steps per block           & 10                    & 10                    & 10                     \\ \hline
Equilibration time (in blocks) & 100                   & 1000                  & 3000                   \\ \hline
Time step $\Delta\tau$ (a.u.)               & $0.005$                 & $0.005$                 & $0.005$                  \\ 
\end{tabular}
\caption{Parameters used in the AFQMC calculations. Parameters not reported here are set to the default setting of ipie \cite{malone2022a}.\label{tab:comp_details}}
\end{table}

\setbox0\hbox{\tabular{@{}l}cc-pVDZ\endtabular}
\setbox1\hbox{\tabular{@{}l}cc-pVTZ\endtabular}
\setbox2\hbox{\tabular{@{}l}cc-pVQZ/def2-QZVPP\endtabular}
\setbox3\hbox{\tabular{@{}l}CBS\endtabular}
\setlength{\extrarowheight}{6pt}

\begin{table}[]
\centering 
\begin{tabular}{p{8mm}p{30mm}p{25mm}p{25mm}p{25mm}p{25mm}p{25mm}p{0mm}p{0mm}}
\hline
Basis & Method &  $E(\ce{^3O_2})$ [Ha] & $E(\ce{^1O_2})$ [Ha] & $E(\ce{O_3})$ [Ha] & $\Delta E (\ce{^1O_2})$ [kJ/mol] & $\Delta E (\ce{O_3})$ [kJ/mol]       &  &   \\
\hline\hline
& Experiment \cite{krupenie1972spectrum, weast1983} & - & - & - & 94.7 & 143.2 \\
\hline\hline
\multirow{6}{*}{\rotatebox{90}{\usebox0}}
& R(O)HF & -149.6083 & -149.5414 & -224.2657 & 175.7 & 385.3 & &  \\ \cline{2-9}
& CCSD(T) & -149.9892 & -149.9405 & -224.9149 & 127.9 & 180.9 & &  \\ \cline{2-9}
& CASSCF & -149.7087 & -149.6757 & -224.4976 & 86.7 & 171.9 & &  \\ \cline{2-9}
& NEVPT2 & -149.9579 & -149.9209 & -224.8759 & 96.9 & 159.8 & & \\ \cline{2-9}
& AFQMC(HF) & -149.9827(7) & -149.9403(6) & -224.9116(14) & 111.2(25) & 164.0(41) & &  \\ \cline{2-9}
& AFQMC(CAS) & -149.9912(3) & -149.9500(2) & -224.9188(4) & 108.0(10) & 178.3(13) & &  \\ \hline
\hline

\multirow{6}{*}{\rotatebox{90}{\usebox1}}
& R(O)HF & -149.6528 & -149.5877 & -224.3405 & 171.1 & 364.4 & &  \\ \cline{2-9}
& CCSD(T) & -150.1536 & -150.1058 & -225.1700 & 125.4 & 158.5 & &  \\ \cline{2-9}
& CASSCF & -149.7519 & -149.7192 & -224.5688 & 86.1 & 155.2 & &  \\ \cline{2-9}
& NEVPT2 & -150.1138 & -150.0782 & -225.1179 & 93.5 & 138.8 & & \\ \cline{2-9}
& AFQMC(HF) & -150.1494(7) & -150.1055(8) & -225.1651(10) & 115.3(28) & 155.1(31) & &  \\ \cline{2-9}
& AFQMC(CAS) & -150.1554(4) & -150.1146(3) & -225.1740(4) & 106.9(13) & 155.1(15) & &  \\ \hline
\hline
\multirow{8}{*}{\rotatebox{90}{\usebox2}}
& R(O)HF & -149.6643 & -149.5995 & -224.3585 & 170.0 & 362.1 & &  \\ \cline{2-9}
& UPBE & -150.2569 & -150.2431 & -225.3366 & 36.3 & 128.0 & &  \\ \cline{2-9}
& UB3LYP & -150.3376 & -150.3214 & -225.4366 & 42.7 & 183.2 & &  \\ \cline{2-9}
& CCSD(T) & -150.2343 & -150.1871 & -225.2929  & 124.0 & 153.8 & &  \\ \cline{2-9}
& CASSCF & -149.7633 & -149.7306 & -224.5866 & 85.9 & 153.2  & &  \\ \cline{2-9}
& NEVPT2 & -150.1928 & -150.1577 & -224.2378 & 92.0 & 134.8 & & \\ \cline{2-9}
& AFQMC(HF) & -150.2279(7) & -150.1880(7) & -225.2901(12) & 104.7(27) & 135.8(36) & &  \\ \cline{2-9}
& AFQMC(CAS) & -150.2354(4)  & -150.1962(3) & -225.2966(4) & 102.9(13) & 148.4(14) & &  \\ \hline
\hline

\multirow{4}{*}{\rotatebox{90}{\usebox3}}
& CCSD(T) & -150.2879 & -150.2409 & -225.3742 &123.4 &151.3 & &  \\ \cline{2-9}
& CASSCF & -149.7662 & -149.7334 & -224.5911 &86.3 & 152.8 & &  \\ \cline{2-9}
& NEVPT2 & -150.2450 & -150.2102 & -225.3169 & 91.4 & 132.9 & & \\ \cline{2-9}
& AFQMC(CAS) & -152.2884(7) & -150.2502(6) & -225.3776(8) &100.4(24) & 144.4(27) & &  \\ \hline
\hline
\end{tabular}
\caption{Results using the experimental geometries given in (\ref{eq:exp_geo}) for various methods and basis sets together with the extrapolated results to the complete basis set (CBS) limit. For all AFQMC(CAS), NEVPT2, and CASSCF calculations, (8e, 6o) and (12e, 9o) active spaces were used for singlet and triplet molecular oxygen, and ozone, respectively. All other calculations were done in the canonical MO basis (R(O)HF). For ozone the experimental value was corrected by the ZPVEs as described in the main text.}
\label{tab:results_exp_geo}
\end{table}

\begin{table}[]
\centering 
\begin{tabular}{p{30mm}p{25mm}p{25mm}p{25mm}p{25mm}p{25mm}p{0mm}p{0mm}}
\hline
Method &  $E(\ce{^3O_2})$ [Ha] & $E(\ce{^1O_2})$ [Ha] & $E(\ce{O_3})$ [Ha] & $\Delta E (\ce{^1O_2})$ [kJ/mol] & $\Delta E (\ce{O_3})$ [kJ/mol]       &  &   \\
\hline\hline
Experiment \cite{krupenie1972spectrum, weast1983} & - & - & - & 94.7 & 143.2 \\
\hline\hline
\multirow{7}{*}{}
R(O)HF & -149.6649 & -149.6018 & -224.3578 & 165.6  & 366.4 & &  \\ \cline{1-8}
UPBE & -150.2568  & -150.2428 & -225.3366 & 36.9 & 127.6 & &  \\ \cline{1-8}
UB3LYP & -150.3377 & -150.3215 & -225.4366 & 42.4 & 183.5 & &  \\ \cline{1-8}
CCSD(T) & -150.2344 & -150.1869 & -225.2929 & 124.7 & 154.0 & &  \\ \cline{1-8}
CASSCF & -149.7632 & -149.7301 & -224.5867 & 87.0 & 152.8 & &  \\ \cline{1-8}
NEVPT2 & -150.1928 & -150.1576 & -225.2378 & 92.4 & 134.7 & & \\ \cline{1-8}
AFQMC(CAS)& -150.2356(4)  & -150.1957(3) & -225.2974(4) & 104.9(12) & 147.1(14) & &  \\ \hline
\hline

\end{tabular}
\caption{Results using the DFT-optimized geometries given in (\ref{eq:dft_geo}) for various methods using the def2-QZVPP basis set for DFT (B3LYP and PBE) and cc-pVQZ otherwise. For all AFQMC(CAS), NEVPT2, and CASSCF calculations, (8e, 6o) and (12e, 9o) active spaces were used for singlet and triplet molecular oxygen, and ozone, respectively. All other calculations were done in the canonical MO basis (R(O)HF). For ozone, the experimental value was corrected by the ZPVEs as described in the main text.}
\label{tab:results_dft_geo}
\end{table}

\end{document}